\documentclass[pre,preprint,showpacs,eqsecnum]{revtex4}
\usepackage{graphicx}
\begin{document}
\title{Directed transport in periodically rocked random sawtooth potentials}
\author{S.~I.~Denisov$^{1,2}$}
\email{stdenis@pks.mpg.de}
\author{T.~V.~Lyutyy$^{2}$}
\author{E.~S.~Denisova$^{2}$}
\author{P.~H\"{a}nggi$^{3,4}$}
\author{H.~Kantz$^{1}$}
\affiliation{$^{1}$Max-Planck-Institut f\"{u}r Physik komplexer Systeme,
N\"{o}thnitzer Stra{\ss}e 38, D-01187 Dresden, Germany\\
$^{2}$Sumy State University, 2 Rimsky-Korsakov Street, UA-40007 Sumy, Ukraine\\
$^{3}$Institut f\"{u}r Physik, Universit\"{a}t Augsburg,
Universit\"{a}tsstra{\ss}e 1, D-86135 Augsburg, Germany \\
$^4$ Department of Physics and Centre for Computational Science and
Engineering, National University of Singapore, Singapore 117542,
Republic of Singapore
}


\begin{abstract}
We study directed transport of overdamped particles in a periodically rocked
random sawtooth potential. Two transport regimes can be identified which are
characterized by a nonzero value of the average velocity of particles and a
zero value, respectively. The properties of directed transport in these regimes
are investigated both analytically and numerically in terms of a random
sawtooth potential and a periodically varying driving force. Precise conditions
for the occurrence of transition between these two transport regimes are
derived and analyzed in detail.
\end{abstract}

\maketitle

\section{INTRODUCTION}

Directed transport in ratchet systems, i.e., devices rectifying undirected
driving forces (both random and deterministic) into directed motion of the
transported particles or localized structures, is much in the limelight of
present activities. The reason is that the rectifying property of these
systems, the so-called ratchet effect, constitutes a theoretical basis for
operating Brownian motors \cite{HM09,HMN,AH,Rei, JAP} which, for example, can
be employed to do surface smoothing \cite{KMWD,DLB, GM,SGMH} or the separation
of particles \cite{BHK, DHR, JKH,Mat, SAM,KSP}. The topic is also closely
related to a variety of yet other intriguing noise-induced transport phenomena
\cite{HM09,HMN,AH, Rei,Linke}.

Usually the action of ratchet systems on the transported particles is described
by a spatially asymmetric, \textit{periodic} potential. The assumption of
strict periodicity of a ratchet potential is technically convenient, but in
many systems the validity of this assumption is not guaranteed. Therefore, if a
sizable spatial aperiodicity is the result of quenched disorder, it is
advantageous to use a \textit{random} ratchet potential for describing the
transport properties in such systems. Within this approach, some of these
properties that result from quenched disorder have already been studied
\cite{HL,March, PASF,AFS, GLZH,ZLAF}.

If the noise arising from the environment can be neglected then the average
drift velocity of overdamped particles in a periodic ratchet potential exhibits
a threshold dependence on the amplitude of the time-periodic driving force (see
also Refs.~\cite{DDH, CPAMH}). In this case only one transport regime with a
nonzero average velocity exists whenever the amplitude of the driving force
exceeds the threshold value. If the amplitude is less than threshold, the
particles remain localized mainly within one period of the ratchet potential,
although co-existing bounded solutions may exist \cite{BHK}. However, this
picture can be changed drastically in ratchet systems containing quenched
disorder. Indeed, if for a random ratchet potential the threshold amplitude
exists and the amplitude of the driving force exceeds this threshold value then
the ordinary transport regime, i.e., transport with a \textit{nonzero} average
velocity of particles and an \textit{arbitrary large} transport distance,
emerges. But if the driving amplitude is somewhat smaller than the threshold
value then a \textit{new transport regime} characterized by a \textit{zero}
average velocity and a \textit{finite} transport distance is expected to be
realized. Moreover, since at the threshold amplitude these regimes merge, it is
plausible that the transport distance approaches infinity if the driving
amplitude tends to the threshold one. The aim of this paper is to study
analytically and numerically these different transport regimes in ratchet
systems with quenched disorder described by a random sawtooth potential.

The paper is structured as follows. In Sec.~II, we introduce the overdamped
equation of motion for particles in a random sawtooth potential driven by a
dichotomously alternating force and formulate the main definitions and
assumptions. Directed transport of particles with a nonzero average velocity is
considered in Sec.~III. Here we derive an explicit formula for the average
velocity in the adiabatic limit and numerically study this transport regime
depending on the amplitude and period of the driving force. In Sec.~IV, we
consider some aspects of directed transport of particles with a vanishing
average velocity. Specifically, we derive the average transport distance of
particles in the preferential direction and study the transition between the
transport regimes with zero and nonzero average velocities. Our main findings
are summarized in Sec~V.

\section{DEFINITIONS AND BASIC EQUATIONS}

We study the directed transport of particles governed by the dimensionless
overdamped equation of motion
\begin{equation}
    \dot{X}_{t} = g(X_{t}) + f(t).
    \label{eqmot}
\end{equation}
Here, $X_{t}$ ($X_{0} = 0$) denotes the particle coordinate, $f(t)$ is a
periodically varying driving force of a period $2T$, and $g(x) = -dU(x) / dx =
\pm g_{\pm}$ presents a dichotomous random force which is generated by a random
sawtooth potential $U(x)$, i.e., a piecewise linear random potential such as
the one depicted in Fig.~1. This random potential $U(x)$ is characterized by
(i) statistically independent random intervals of lengths $s_{j}$ which are
distributed with the probability densities $p_{+}(s)$ and $p_{-}(s)$ for even
numbered ($j=2n$, $n=0,\pm 1,\ldots$) and odd numbered ($j=2n+1$) intervals,
respectively, (ii) two deterministic slopes $-g_{+}$ and $g_{-}$ ($g_{+}>
g_{-}>0$), and (iii) the constraint condition $g_{+}s_{+} = g_{-}s_{-}$ with
$s_{\pm}= \int_{0}^{\infty} ds\, sp_{\pm}(s)<\infty$ denoting the average
lengths of even, $s_{+}$, and odd, $s_{-}$, intervals. The last condition
implies that the average value of the dichotomous random force equals zero,
i.e.,
\begin{eqnarray}
    \lim_{L \to \infty} \frac{1}{2L}\int_{-L}^{L}dx\,g(x)  \!\!&=&\!\!
    \lim_{L \to \infty} \bigg(g_{+}\frac{L_{+}}{L} - g_{-}\frac{L_{-}}{L}
    \bigg)
    \nonumber\\[6pt]
    \!\!&=& \!\! g_{+}s_{+} - g_{-}s_{-} = 0,
    \label{AvFor1}
\end{eqnarray}
where $2L_{+}$ and $2L_{-}$ denote the total lengths of the even and the odd
intervals on the interval $(-L,L)$. In addition, we assume that at the origin
of the coordinate system all sample paths of this potential change its slope
from $-g_{+}$ (at $x=-0$) to $g_{-}$ (at $x=+0$), i.e., $g(\pm 0) = \mp
g_{\mp}$. We also note that the special case of the above defined potential,
when $g_{+}=g_{-}$, has been invoked to study the statistical properties of the
arrival time and the arrival position of particles in a medium with quenched
dichotomous disorder driven by a constant force \cite{DKDH1, DKDH2}.

This chosen form of the random potential in Fig.~1 possesses random (i.e.,
non-periodic) barrier widths with corresponding (random) barrier heights and
comprises the complexity of more general realizations of random landscapes
while at the same time allowing for an explicit analytical treatment (see also
the further remarks given before Sec. V below).

The periodic driving  force $f(t)$ is assumed to be alternating, i.e., $f(t)=
(-1)^{k+1}f$, where $f\,(>0)$ is the amplitude of $f(t)$ or the driving
strength, $k=[t/T]+1$, and $[t/T]$ is the integer part of $t/T$. This force can
therefore be explicitly written as
\begin{equation}
    f(t) = \left\{ \begin{array}{ll}
    f, \quad (2k-2)T \leq t < (2k-1)T \\ [6pt]
    -f, \quad (2k-1)T \leq t < 2kT
    \end{array}
    \right.
    \label{f(t)}
\end{equation}
($k=1,2,\ldots$) and, since
\begin{equation}
    \lim_{\tau \to \infty} \frac{1}{\tau}\int_{0}^{\tau}dt\,f(t) = 0,
    \label{AvFor2}
\end{equation}
its average value equals zero as well.

In accordance with Eq.~(\ref{eqmot}), if $f\leq g_{-}$ then the particle
remains localized in the initial state for all $t>0$; the particle coordinate
depends on time only if $f>g_{-}$. Although the average values of the forces
$g(x)$ and $f(t)$ are zero, a systematic displacement of particles along the
positive direction (since $g_{+}>g_{-}$) of the axis $x$ may nevertheless
exist. In other words, the potential $U(x)$ can rectify the periodic motion of
particles induced by the periodic alternating force $f(t)$ into directed
transport. This phenomenon is known as the ratchet effect \cite{HMN,AH,Rei,
JAP}.

\section{AVERAGE TRANSPORT VELOCITY}

\subsection{Long-period limit}

We next define the average transport velocity $v_{T}$ of particles in the
following way:
\begin{equation}
    v_{T} = \lim_{t\to\infty} \frac{\langle X_{t} \rangle}{t},
    \label{defv1}
\end{equation}
where the angular brackets denote an averaging over the sample paths of $g(x)$.
In the case of long-lasting period of the driving force, $T\to\infty$, it is
possible to derive the exact formula for $v_{\infty}$. The starting point is
that in the definition (\ref{defv1}) we can replace $t$ by $2T$ and $X_{2T}$ by
$X_{T}-Y_{T}$, yielding
\begin{equation}
    v_{\infty} = \lim_{T\to\infty} \frac{\langle X_{T} \rangle -
    \langle Y_{T} \rangle}{2T},
    \label{defv2}
\end{equation}
where $X_{T}$ is the displacement of the particle during the first half-period
of $f(t)$, and $Y_{T} = X_{T} - X_{2T}$ is the displacement during the second
half-period. As pointed out above, at $f \leq g_{-}$ the particle stays in the
initial state for all $t$, and so $v_{\infty}=0$. A simple analysis of
Eq.~(\ref{eqmot}) shows that if $f>g_{-}$ then $X_{T}>0$ and $\lim_{T
\to\infty} \langle X_{T} \rangle/T \in [f-g_{-},f+g_{+}]$, i.e., this limit is
always nonvanishing. In contrast, the sign of $Y_{T}$ depends on $f$ and
$g(X_{T})$, and the limit $\lim_{T\to\infty} \langle Y_{T} \rangle/T$ can
either be zero or nonzero. Specifically, if $g_{-}<f\leq g_{+}$ then $Y_{T}>0$
at $g(X_{T}) = -g_{-}$ and $Y_{T}\leq 0$ at $g(X_{T}) = g_{+}$. Since in this
case the particle moves during the second half-period only in one odd interval
(if $g(X_{T}) = -g_{-}$) or in one even interval (if $g(X_{T}) = g_{+}$), we
obtain $\lim_{T\to\infty} \langle Y_{T} \rangle/T =0$. On the contrary, if $f>
g_{+}$ then $Y_{T}>0$ and the particle passes during the second half-period an
infinite number of intervals $s_{j}$ when $T\to \infty$. In this case $\lim_{T
\to\infty} \langle Y_{T} \rangle/T \in [f-g_{+},f+g_{-}]$, i.e., the limit is
nonzero. One therefore expects that the dependence of $v_{\infty}$ on $f$
differs in these two cases.

Let us first derive the average velocity for $g_{-}<f\leq g_{+}$. In this case
$\lim_{T\to\infty} \langle Y_{T} \rangle/T =0$ and Eq.~(\ref{defv2}) becomes
\begin{equation}
    v_{\infty} = \lim_{T\to\infty} \frac{\langle X_{T} \rangle}{2T}.
    \label{defv3}
\end{equation}
Representing $X_{T}$ in the form $X_{T} = X_{T}^{+} + X_{T}^{-}$,
where $X_{T}^{+}$ and $X_{T}^{-}$ are the total lengths of the even
and odd intervals $s_{j}$ on the interval $(0,X_{T})$, respectively,
and using the relation $g_{+}\langle X_{T}^{+} \rangle =
g_{-}\langle X_{T}^{-} \rangle$, which follows from the condition
$g_{+} s_{+} = g_{-} s_{-}$, we find
\begin{equation}
    \langle X_{T} \rangle = \langle X_{T}^{-} \rangle\left( 1 +
    \frac{g_{-}}{g_{+}} \right).
    \label{rel1}
\end{equation}
On the other hand, because the particle passes the even and odd
intervals with the velocities $f+g_{+}$ and $f-g_{-}$, respectively,
we have
\begin{eqnarray}
    T \!\!&=&\!\! \frac{\langle X_{T}^{+} \rangle}{f+g_{+}} +
    \frac{\langle X_{T}^{-} \rangle}{f-g_{-}}
    \nonumber\\[6pt]
    \!\!&=& \!\! \langle X_{T}^{-} \rangle\left( 1 + \frac{g_{-}}{g_{+}}
    \right) \frac{f+g_{+}-g_{-}}{(f+g_{+})(f-g_{-})}.
    \label{rel2}
\end{eqnarray}
Finally, substituting Eqs.~(\ref{rel1}) and (\ref{rel2}) into
Eq.~(\ref{defv3}), the resulting  average velocity assumes the form
\begin{equation}
    v_{\infty} = \frac{(f+g_{+})(f-g_{-})}{2(f+g_{+}-g_{-})}.
    \label{v1}
\end{equation}

If $f>g_{+}$ then the average velocity of particles is defined by
Eq.~(\ref{defv2}). Introducing by analogy with the previous case the total
lengths $Y_{T}^{+}$ and $Y_{T}^{-}$ of the even and odd intervals $s_{j}$ on
the interval $(0,Y_{T})$, for the second half-period of $f(t)$ we obtain
\begin{equation}
    \langle Y_{T} \rangle = \langle Y_{T}^{-} \rangle \left( 1 +
    \frac{g_{-}}{g_{+}} \right)
    \label{rel3}
\end{equation}
and, likewise,
\begin{eqnarray}
    T \!\!&=&\!\! \frac{\langle Y_{T}^{+} \rangle}{f-g_{+}} +
    \frac{\langle Y_{T}^{-} \rangle}{f+g_{-}}
    \nonumber\\[6pt]
    \!\!&=& \!\! \langle Y_{T}^{-} \rangle\left( 1 + \frac{g_{-}}
    {g_{+}} \right)\frac{f-g_{+}+g_{-}}{(f-g_{+})(f+g_{-})}.
    \label{rel4}
\end{eqnarray}
Calculating with the help of these formulas the limit $\lim_{T\to\infty}
\langle Y_{T} \rangle/T$, Eq.~(\ref{defv2}) gives the result
\begin{equation}
    v_{\infty} = \frac{(f+g_{+})(f-g_{-})}{2(f+g_{+}-g_{-})}
    -\frac{(f-g_{+})(f+g_{-})}{2(f-g_{+}+g_{-})},
    \label{v2a}
\end{equation}
which can be simplified to read
\begin{equation}
    v_{\infty} = \frac{g_{+}g_{-}(g_{+}-g_{-})}{f^{2}-(g_{+}-g_{-})^{2}}.
    \label{v2b}
\end{equation}

Thus, in the adiabatic limit the average velocity of particles in a random
sawtooth potential $U(x)$ driven by a periodically alternating force $f(t)$ is
given by Eq.~(\ref{v1}) if $g_{-}<f\leq g_{+}$, and Eq.~(\ref{v2b}) if
$f>g_{+}$. It is important to note that these results do not depend on the
concrete distributions of the intervals $s_{j}$. In fact, in evaluating the
average velocity we used  only the condition that the probability densities
$p_{\pm}(s)$ possess finite first moments $s_{\pm}$.

Since by assumption $g_{+}>g_{-}$, the transport of particles occurs with the
average velocity $v_{\infty}$ in the positive direction of the axis $x$. In
accordance with Eqs.~(\ref{v1}) and (\ref{v2b}), $v_{\infty}$ is a nonmonotonic
function of $f$ which assumes the maximum value
\begin{equation}
    v_{\infty\,\mathrm{max}} = \frac{g_{+}(g_{+}-g_{-})}{2g_{+}-g_{-}}
    \label{vmax}
\end{equation}
for $f=g_{+}$, i.e., at the point where the character of $v_{\infty}$ as a
function of $f$ changes qualitatively. If, on the other hand, $g_{+}<g_{-}$
then the transport of particles occurs in the negative direction of the axis
$x$. In this case, the average velocity (and the corresponding conditions for
$f$) is determined by Eqs.~(\ref{v1}) and (\ref{v2b}) in which $g_{+}$ and
$g_{-}$ must be replaced by $g_{-}$ and $g_{+}$, respectively. At $g_{+}=g_{-}$
the average velocity $v_{\infty}$ equals zero for all $f$.

The predicted  dependence of $v_{\infty}$ on $f$ and the numerical results
obtained by simulation of Eq.~(\ref{eqmot}) are depicted in Fig.~2 for the case
with $g_{+}=6$ and $g_{-}=2$. For each $f$ the numerical average velocity,
\begin{equation}
    v_{\mathrm{sim}} = \frac{1}{N}\sum_{i=1}^{N} \frac{X_{2T}^{(i)}}{2T},
    \label{vsim}
\end{equation}
shown in this figure by the triangular symbols, was calculated for $N=2\times
10^{2}$ runs of the motion equation (\ref{eqmot}). Before each run, a new
realization of the dichotomous force $g(x)$ was generated in accordance with
the uniform probability densities
\begin{equation}
    p_{\pm}(s) = \left\{ \begin{array}{ll}
    (2d_{\pm})^{-1}, \quad 0 \leq s \leq 2d_{\pm} \\ [6pt]
    0, \quad s > 2d_{\pm}
    \end{array}
    \right..
    \label{uniform}
\end{equation}
Since in this case $s_{\pm}=d_{\pm}$, we chose $d_{+}=0.5$ and $d_{-}=1.5$ in
order to satisfy the condition $g_{+}s_{+}=g_{-}s_{-}$. Finally, to ensure that
the particle passes a large number of the intervals $s_{j}$, we chose $T=10^{2}
d_{-}/(f-g_{-})$ if $f\in (g_{-},g_{+}]$, and $T=10^{2}\max{ d_{\pm}/(f-
g_{\pm})}$ if $f>g_{+}$. As seen from Fig.~2, our numerical results obtained in
such a way are in excellent agreement with theory. It should be noted, however,
that we used the uniform distributions for the intervals $s_{j}$ only for
illustrative purposes: The average velocity $v_{T}$ of particles does not
depend on $p_{\pm}(s)$ in the limit $T\to\infty$.

\subsection{Nonadiabatic driving}

In contrast to the previous case, if $T$ is finite then the average velocity
$v_{T}$ depends on the explicit form of the probability densities $p_{\pm}(s)$.
It is important to note that in this case the condition $f>g_{-}$ does not
guarantee that $v_{T}> 0$ (at $g_{+}>g_{-}$). Specifically, if $\int_{a}
^{\infty}ds\,p_{-}(s)> 0$ for arbitrary large (but finite) $a$, i.e., if
$p_{-}(s)$ is the probability density with unbounded support, then $\langle
X_{\infty} \rangle < \infty$ and so $v_{T}=0$ for all finite $f$ and $T$.
However, even if $p_{-}(s)=0$ for $s>b$, i.e., if $p_{-}(s)$ has bounded
support, the average velocity equals zero as well if $T< T_{\mathrm{th}} =
b/(f- g_{-})$, where $T_{\mathrm{th}}$ is the threshold half-period. From a
physical point of view, the condition $v_{T}=0$, i.e., the absence of directed
transport of particles towards infinity, arises from the fact that there is a
nonzero probability of those (odd) intervals $s_{j}$ that cannot be overcome by
particles during a positive pulse of $f(t)$. The existence of directed
transport of particles with a zero average velocity and a finite transport
distance will be considered in more detail in the next section.

In accordance with the above discussion, directed transport of particles with a
nonzero average velocity exists only if both conditions, $f>g_{-}$ and $T>
T_{\mathrm{th}}$, hold true. Since the latter condition is more restrictive
than the former, it is the latter that determines the criterion of directed
transport of particles with nonzero average velocity $v_{T}$. In particular, if
$p_{-}(s)$ is the uniform probability density, see Eq.~(\ref{uniform}), then
this criterion can be written in the form
\begin{equation}
    f > f_{\mathrm{th}} = g_{-} + \frac{2d_{-}}{T},
    \label{cond}
\end{equation}
where $f_{\mathrm{th}}$ is the threshold amplitude of the driving force $f(t)$.
The dependencies of $v_{T}$ on $f$ obtained by the numerical simulation of the
motion equation (\ref{eqmot}) are shown in Fig.~3. Typical solutions of this
equation for $f\in (f_{\mathrm{th}},g_{+})$ and $f>g_{+}$ are illustrated in
Fig.~4.

\section{DIRECTED TRANSPORT WITH ZERO AVERAGE VELOCITY}

In the case of zero average velocity the particles cannot be transported to an
arbitrary large distance along the axis $x$ in a common way. Instead, for each
realization of $g(x)$ the particles are transported to any position in whose
vicinity they oscillate (see Fig.~5). For random $g(x)$ these positions are
random as well and, since $g_{+}>g_{-}$, they are preferably distributed at
$x>0$. Our next objective is to find the average distance $\langle l \rangle$
from the origin of the coordinate system to these positions in the positive
direction of the axis $x$.

Let us assume that the interval $s_{2n+1}$ with some $n\,(\geq 0)$ is the
\textit{first} odd interval satisfying the condition $s_{2n+1}>\Delta$, where
\begin{equation}
\Delta = (f-g_{-})T
\end{equation}
is the minimal displacement of particles during a positive pulse of the driving
force $f(t)$. In other words, $s_{2n+1}$ is the first interval which is not
crossed by particles in the positive direction of the axis $x$. The distance
from the coordinate origin to this interval is given by
\begin{equation}
    l_{2n} = \sum_{j=1}^{2n}s_{j}
    \label{dist}
\end{equation}
if $n\geq 1$, and $l_{0}=0$ if $n=0$. We also introduce the probability
\begin{equation}
    w = 1 - \int_{\Delta}^{\infty}ds\,p_{-}(s) =
    \int_{0}^{\Delta}ds\,p_{-}(s)
    \label{prob}
\end{equation}
that the length of the odd interval is smaller than $\Delta$. In accordance
with these definitions, the probability density $P(l)$ that $l=l_{2n}$ (with
$n\geq 0$) can be written in the form
\begin{eqnarray}
    P(l) \!\!&=&\!\! (1-w)\sum_{n=1}^{\infty}\int_{0}^{\Delta}\!\!
    \ldots\!\int_{0}^{\Delta}\bigg(\prod_{j=1}^{n}ds_{2j-1}p_{-}
    (s_{2j-1})\bigg)
    \nonumber\\[6pt]
    && \!\! \times \int_{0}^{\infty}\!\!\ldots\!\int_{0}^{\infty}
    \bigg(\prod_{k=1}^{n}ds_{2k}p_{+}(s_{2k})\bigg)\delta(l- l_{2n})
    \nonumber\\[6pt]
    && \!\! + \,(1-w)\,\delta(l),
    \label{P(l)}
\end{eqnarray}
where $\delta(\cdot)$ is the Dirac $\delta$ function. Using the geometric
series formula $\sum_{n=0}^{\infty}w^{n} = (1-w)^{-1}$, it is not difficult to
verify that $P(l)$ is properly normalized, i.e.,
\begin{eqnarray}
    \int_{0}^{\infty}dl\,P(l) \!\!&=&\!\! (1-w)\sum_{n=1}^{\infty}
    w^{n} +1-w
    \nonumber\\[4pt]
    &=&\!\! (1-w)\sum_{n=0}^{\infty}w^{n} = 1.
    \label{norm}
\end{eqnarray}

The average transport distance, i.e., the mean value of $l_{2n}$, is defined in
the usual way:
\begin{equation}
    \langle l \rangle = \int_{0}^{\infty}dl\,lP(l).
    \label{defl}
\end{equation}
In order to calculate this integral, we first note that, according to
Eq.~(\ref{dist}),
\begin{equation}
    \int_{0}^{\infty}dl\,l \delta(l-l_{2n}) = \sum_{m=1}^{n}s_{2m-1} +
    \sum_{m=1}^{n}s_{2m}.
    \label{sum1}
\end{equation}
Then, substituting the probability density (\ref{P(l)}) into (\ref{defl}) and
taking into account the formulas
\begin{equation}
    \int_{0}^{\Delta}\!\!\ldots\!\int_{0}^{\Delta}\!\bigg(
    \prod_{j=1}^{n}ds_{2j-1}p_{-}(s_{2j-1})\!\bigg)\!\sum_{m=1}^
    {n}\! s_{2m-1} = n\tilde{s}_{-}w^{n-1},
    \label{int1}
\end{equation}
where $\tilde{s}_{-} = \int_{0}^{\Delta}ds\,sp_{-}(s)$, and
\begin{equation}
    \int_{0}^{\infty}\!\!\ldots\!\int_{0}^{\infty}\!\bigg(
    \prod_{k=1}^{n}ds_{2k}p_{+}(s_{2k})\bigg)\!\sum_{m=1}^
    {n}s_{2m} = ns_{+}\;,
    \label{int2}
\end{equation}
the expression (\ref{defl}) can be reduced to read
\begin{equation}
    \langle l \rangle = (1-w)(\tilde{s}_{-} + s_{+}w)
    \sum_{n=1}^{\infty}nw^{n-1}.
    \label{av1}
\end{equation}
Finally, using the formula $\sum_{n=1}^{\infty}nw^{n-1} = (1-w)^{-2}$, we
obtain for the average transport distance the following remarkably simple
result:
\begin{equation}
    \langle l \rangle = \frac{\tilde{s}_{-} + s_{+}w}{1-w}.
    \label{av2}
\end{equation}

It is important to note that Eq.~(\ref{av2}) represents the average distance to
the first impassable interval in the \textit{positive} direction of the axis
$x$, i.e., the average value of the maximum displacement of particles in the
preferred direction. If $f\in(g_{-},g_{+})$ then $X_{t}\geq 0$ for all sample
paths of $g(x)$ and thus the average displacement of particles,
$\lim_{\tau\to\infty} (1/ \tau) \int_{0}^{\tau}dtX_{t}$, relates closely to
$\langle l \rangle$. But when $f>g_{+}$ then there exists a set of sample
paths, whose total probability is nonzero, on which the particles are
transported in the negative direction of the axis $x$. As a consequence, in
this case the average displacement of particles is, in general, smaller than
$\langle l \rangle$.

According to Eq.~(\ref{av2}), the average distance $\langle l \rangle$ is
finite if $w\neq 1$. If the probability density $p_{-}(s)$ has unbounded
support then this condition holds for \textit{all} half-periods $T$ of the
driving force $f(t)$. Otherwise, i.e., in the case of bounded support, $\langle
l \rangle$ may be finite or infinite depending on the value of $T$. In order to
illustrate the distinctive features of directed transport in these two cases,
we next calculate $\langle l \rangle$ for the exponential and uniform
probability densities $p_{\pm}(s)$, which represent the probability densities
with unbounded and bounded support, respectively.

\subsection{Exponentially distributed intervals}

For the exponential probability densities
\begin{equation}
    p_{\pm}(s) = \lambda_{ \pm} e^{-\lambda_{\pm} s},
    \label{exp}
\end{equation}
where $\lambda_{ \pm}$ are the rate parameters, we have $s_{\pm} = \lambda_{
\pm}^{-1}$, $w = 1 - e^{-\lambda_{-}\Delta}$, and
\begin{equation}
    \tilde{s}_{-} = \frac{1}{\lambda_{-}}(1 - e^{-\lambda_{-}\Delta} -
    \lambda_{-}\Delta \, e^{-\lambda_{-}\Delta})\;.
    \label{tilds}
\end{equation}
Therefore, in this case the formula (\ref{av2}) becomes
\begin{equation}
    \langle l \rangle = \frac{\lambda_{-} + \lambda_{+}}
    {\lambda_{-}\lambda_{+}}(e^{\lambda_{-}\Delta} - 1) - \Delta\,.
    \label{av3}
\end{equation}
For given $g_{+}$ and $g_{-}$, the parameters $\lambda_{\pm}$ and $g_{\pm}$ are
not independent because, in accordance with Eq.~(\ref{AvFor1}), the condition
$g_{+} \lambda_{-} = g_{-} \lambda_{+}$ must hold. Eliminating with the help of
this relation the parameter $\lambda_{+}$, Eq.~(\ref{av3}) yields
\begin{equation}
    \langle l \rangle = \frac{1}{\lambda_{-}}\bigg( 1 + \frac{g_{-}}
    {g_{+}} \bigg)(e^{\lambda_{-}\Delta} - 1) - \Delta\,.
    \label{av4}
\end{equation}

According to this result, the average distance $\langle l \rangle$ is finite,
and so $v_{T}=0$, for all finite $f>g_{-}$ and $T$. In other words, in the case
of exponential distributions of the interval $s_{j}$ the directed transport of
particles always occurs with zero average velocity. This feature of directed
transport arises from the fact that the probability density $p_{-}(s)$ has
unbounded support. As it follows from Eq.~(\ref{av4}), the average distance
grows linearly with $\Delta$, $\langle l \rangle = (g_{-}/g_{+}) \Delta$, if
$\lambda_{-} \Delta \ll 1$, and exponentially, $\langle l \rangle =
\lambda_{-}^{-1} (1 + g_{-}/g_{+}) e^{\lambda_{-} \Delta}$, if $\lambda_{-}
\Delta \gg 1$. Our analytical results are in full agreement with the numerical
simulations (see Fig.~6).

\subsection{Uniformly distributed intervals}

If the intervals $s_{j}$ are distributed with uniform probability densities
(\ref{uniform}) then $s_{\pm} = d_{\pm}$,
\begin{equation}
    w = \left\{ \begin{array}{ll}
    \Delta (2d_{-})^{-1}, \quad 0 < \Delta < 2d_{-} \\ [6pt]
    1, \quad \Delta \geq 2d_{-}
    \end{array}
    \right.,
    \label{w}
\end{equation}
and
\begin{equation}
    \tilde{s}_{-} = \left\{ \begin{array}{ll}
    \Delta^{2} (4d_{-})^{-1}, \quad 0 < \Delta < 2d_{-} \\ [6pt]
    d_{-}, \quad \Delta \geq 2d_{-}
    \end{array}
    \right..
    \label{s}
\end{equation}
According to these results, Eq.~(\ref{av2}) for $\Delta \geq 2d_{-}$, i.e.,
$f\geq f_{\mathrm{th}}$, yields $\langle l \rangle = \infty$. In contrast, if
$0 < \Delta < 2d_{-}$, i.e., $f\in(g_{-},f_{\mathrm{th}})$, then
Eq.~(\ref{av2}) reduces to
\begin{equation}
    \langle l \rangle = \frac{\Delta(2d_{+} + \Delta)}
    {2(2d_{-} - \Delta)}.
    \label{av5}
\end{equation}
Since $g_{+}d_{+} = g_{-}d_{-}$, the last formula can be rewritten in the form
\begin{equation}
    \langle l \rangle = d_{-}\bigg(1+ \frac{g_{-}}{g_{+}}\bigg)
    \frac{f-g_{-}}{f_{\mathrm{th}} - f} - \frac{T}{2}(f-g_{-}).
    \label{av6}
\end{equation}

Thus, depending on $f$, two regimes of directed transport exist. The first
occurs at $f\in(g_{-},f_{\mathrm{th}})$ and is characterized by a zero average
velocity $v_{T}$ and a finite transport distance (\ref{av6}). The second, with
a nonzero $v_{T}$ and an infinite $\langle l \rangle$, takes place at $f>
f_{\mathrm{th}}$. At the threshold amplitude $f=f_{\mathrm{th}}$ the transition
between these regimes occurs. Like in the previous case, the dependencies of
the average transport distance $\langle l \rangle$ on $f$, which follow from
(\ref{av6}), are fully corroborated by our numerical simulations (see Fig.~7).

We note that the random sawtooth potentials account for the influence of
quenched disorder in non-periodic ratchet systems and at the same time allow
for a full analytical description of the ratchet effect. In the case of other
random ratchet potentials a rigorous theoretical analysis of directed transport
becomes extremely cumbersome without providing prominent additional insight.
Put differently, the above analysis evidences that qualitatively the same
results hold for a wider class of random ratchet potentials that produce the
random forces $g(x)$ varying in the interval $(-g_{-},g_{+})$ and assuming a
zero mean value. Specifically, if the distances between the nearest global
maxima of $g(x)$ are distributed with unbounded support then only one transport
regime of particles with $v_{T}=0$ can be realized. The reason for this is the
same as in the case of a dichotomous random force: For any finite half-period
$T$ of the driving force $f(t)$ there is always a nonzero probability for
distances that cannot be overcome by particles during a positive pulse of
$f(t)$. Accordingly, if the support is bounded then two transport regimes with
$v_{T}>0$ (when $T$ is sufficiently large) and $v_{T}=0$ (when $T$ is
sufficiently short) exist.

\section{CONCLUSIONS}

We have studied the directed transport of particles in absence of noise which
are driven by a periodically alternating force in a viscous medium with
quenched disorder. The influence of quenched disorder is modeled by a random
sawtooth potential that generates a dichotomous random force with zero mean. We
could show that, depending on the characteristics of the dichotomous and
driving forces, two regimes of directed transport occur, namely, with a nonzero
average velocity and with a vanishing average velocity.

The main result which we have obtained for the former regime is an explicit
formula for the average transport velocity in the long-period limit of the
driving force. An important feature of this limiting formula is that it does
not depend on the probability densities of the intervals characterizing the
dichotomous random force. We have shown numerically that for finite periods of
the driving force the average transport velocity is always less than the
limiting one if all other parameters are kept the same.

In order to  characterize the transport regime with a zero average velocity, we
have calculated analytically the average value of the maximum displacement of
particles in the preferred transport direction. This quantity is finite and so
the average velocity of particles is zero if the probability density of the odd
intervals characterizing the dichotomous force has unbounded support.
Otherwise, i.e., if this probability density has bounded support, the average
velocity can be either zero or nonzero, depending on the characteristics of the
dichotomous and driving forces. We have applied the uniform probability
densities for the quantitative study of the transport properties in these
regimes and for describing the transition between them. All our theoretical
predictions are nicely confirmed by our numerical simulations.

\section*{ACKNOWLEDGMENTS}

S.I.D. acknowledges the support of the EU through Contract No.
MIF1-CT-2006-021533, and P.H. acknowledges financial support by the Deutsche
For\-schungs\-ge\-mein\-schaft via the Collaborative Research Centre SFB-486,
Project No. A10 and by the German Excellence Initiative via the \textit
{Nanosystems Initiative Munich} (NIM).

\newpage

\begin{figure}
    \centering
    \includegraphics[totalheight=6cm]{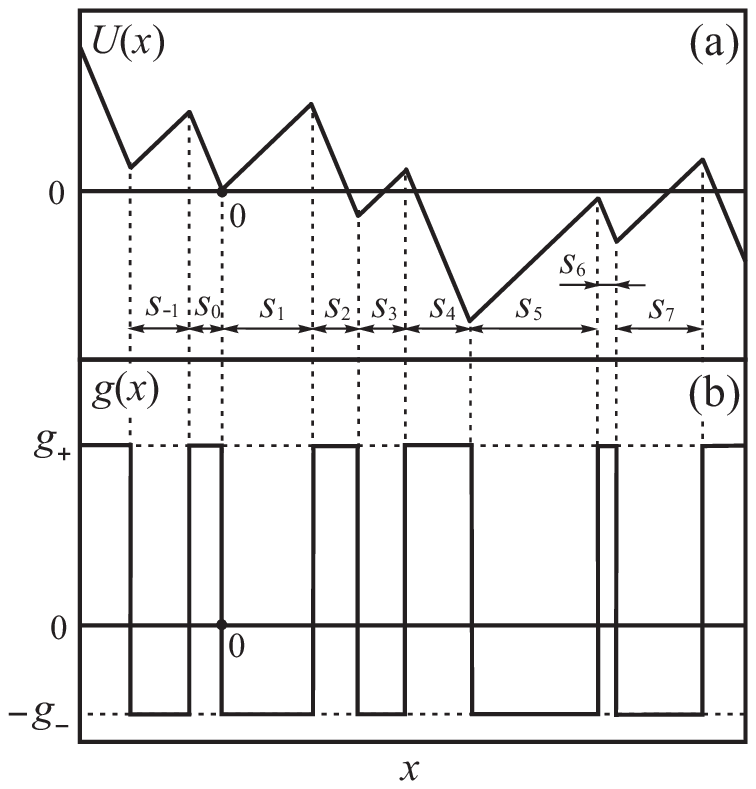}
    \caption{\label{fig1} Schematic representation of the random
    sawtooth potential $U(x)$ (a) and the corresponding dichotomous
    random force $g(x)$ (b) as functions of the spatial coordinate $x$. }
\end{figure}

\begin{figure}
    \centering
    \includegraphics[totalheight=5cm]{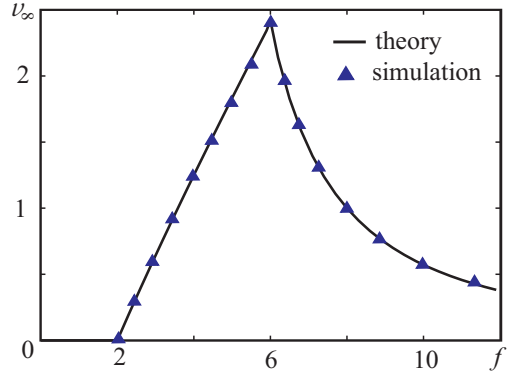}
    \caption{\label{fig2} (Color online) Average transport velocity
    $v_{\infty}$ of particles as a function of the driving strength
    $f$ in the adiabatic limit. The solid lines represent the
    theoretical results obtained from Eqs.~(\ref{v1}) and (\ref{v2b}),
    and the triangular symbols (blue online) indicate results derived
    from the numerical simulations of Eq.~(\ref{eqmot}). The presented
    results correspond to the dichotomous random force $g(x)$ with
    $g_{+}=6$ and $g_{-}=2$. }
\end{figure}

\begin{figure}
    \centering
    \includegraphics[totalheight=5cm]{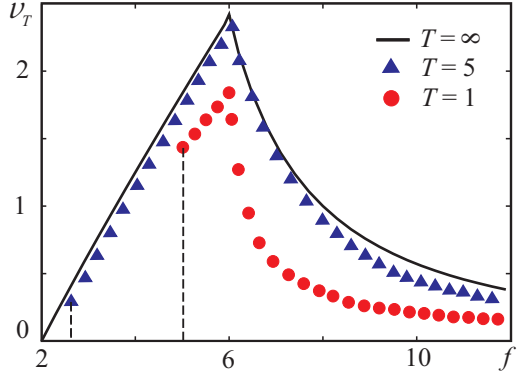}
    \caption{\label{fig3} (Color online) Average transport velocity
    $v_{T}$ as a function of the driving strength $f$ for different
    values of the half-period $T$. The triangular (blue online) and
    circular (red online) symbols represent the numerical results
    obtained via $N=10^{3}$ runs of Eq.~(\ref{eqmot}) and by using
    the numerical average velocity (\ref{vsim}) in which $2T$ is
    replaced by $40T$. The theoretical dependence of $v_{T}$ on $f$
    for $T=\infty$ (solid lines) reproduces the average velocity
    $v_{\infty}$ from Fig.~2 and is shown for comparison only.
    The parameters characterizing the dichotomous random force $g(x)$
    whose intervals $s_{j}$ are distributed with uniform
    probability densities (\ref{vsim}) are chosen to be $g_{+}=6$,
    $g_{-}=2$, $d_{+}=0.5$, and $d_{-}=1.5$. According to
    Eq.~(\ref{cond}), $f_{\mathrm{th}} = 2.6$ for $T=5$ and
    $f_{\mathrm{th}} = 5$ for $T=1$. }
\end{figure}

\begin{figure}
    \centering
    \includegraphics[totalheight=5cm]{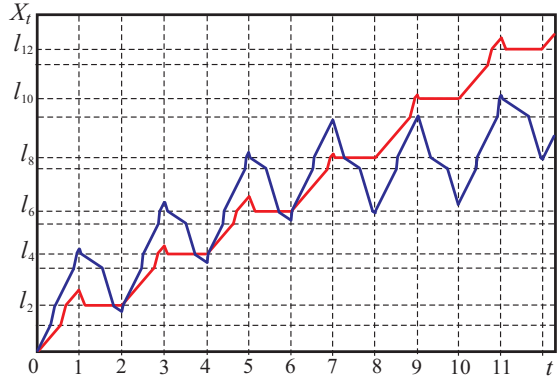}
    \caption{\label{fig4} (Color online) Illustrative realizations
    of the particle coordinate $X_{t}$ in the case of nonzero
    average velocity $v_{T}$. The parameters of the dichotomous
    random force $g(x)$ are chosen as in Fig.~3, $T=1$, and
    $f_{\mathrm{th}} = 5$. The transport regime with $v_{T}>0$
    occurs at $f>f_{\mathrm{th}}$, and $X_{t}$ displays
    different behavior for $f\in (f_{\mathrm{th}}, g_{+})$ and
    $f>g_{+}$. The line with horizontal segments (red online)
    represents $X_{t}$ for $f=5.5$ (in this case $f<g_{+}=6$,
    $v_{T=1}=1.62$), and the other line (blue online) represents
    $X_{t}$ for $f=8$ ($f>g_{+}$, $v_{T=1}=1.00$). For
    convenience, the distances $l_{n} = \sum_{j=1}^{n}s_{j}$,
    which correspond to a given sample path of $g(x)$, are
    shown for even $n$ only.}
\end{figure}

\begin{figure}
    \centering
    \includegraphics[totalheight=5cm]{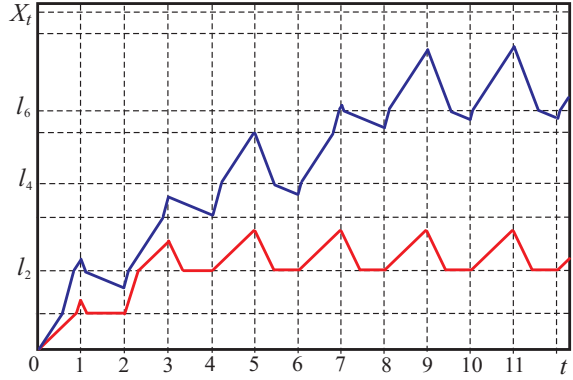}
    \caption{\label{fig5} (Color online) Illustrative realizations
    of the particle coordinate $X_{t}$ in the case of zero average
    velocity $v_{T}$. The parameters of the dichotomous random
    force $g(x)$ are the same as in Fig.~3, $T=0.5$, and $f_{\mathrm{th}}
    =8$. In this case, the transport regime with $v_{T}=0$ occurs only if
    $f\in (g_{-}, f_{\mathrm{th}})$, and $X_{t}$ shows different behavior
    for $f< g_{+}$ and $f> g_{+}$. The lower line (red online) represents
    $X_{t}$ for $f=5.5$ ($f<g_{+}=6$, $\langle l \rangle = 1,93$),
    and the upper line (blue online) for $f=7.5$ ($f>g_{+}$,
    $\langle l \rangle = 20,63$).}
\end{figure}

\begin{figure}
    \centering
    \includegraphics[totalheight=5cm]{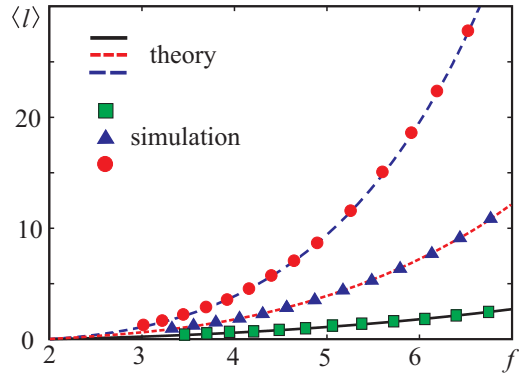}
    \caption{\label{fig6} (Color online) Average transport distance
    $\langle l \rangle$ as a function of the driving strength $f$
    for exponentially distributed intervals $s_{j}$.  The theoretical
    curves are derived from Eq.~(\ref{av4}) with $g_{+}=6$, $g_{-}=2$,
    $\lambda_{-}=1/3$, and $T=0.5$ (solid line), $T=1$ (short-dashed
    line, red online), and $T=1.5$ (long-dashed line, blue online).
    The symbols (in color online) depict the numerical results obtained
    by (i) generating a sample path of $g(x)$ in accordance with
    exponential distributions (\ref{exp}), (ii) finding the distance
    (\ref{dist}) to the first interval $s_{2n+1}$ whose length exceeds
    $\Delta$, and (iii) averaging this distance over $10^{3}$
    realizations of $g(x)$.}
\end{figure}

\begin{figure}
    \centering
    \includegraphics[totalheight=5cm]{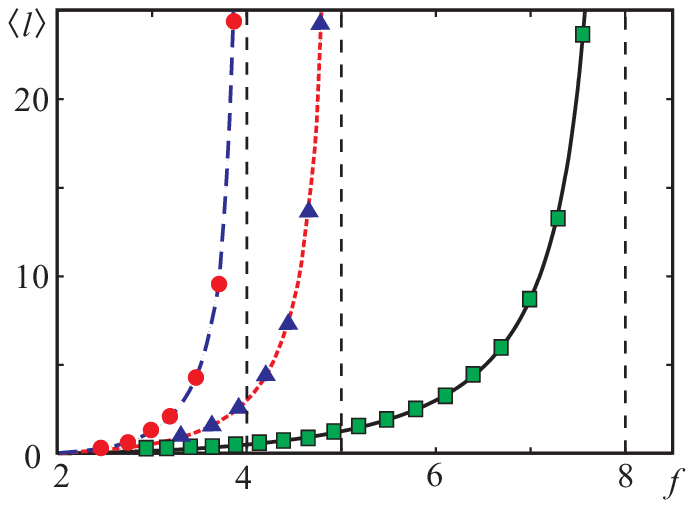}
    \caption{\label{fig7} (Color online) Average transport distance
    $\langle l \rangle$ as a function of the driving strength  $f$
    for uniformly distributed intervals $s_{j}$. The theoretical
    curves are obtained from Eq.~(\ref{av6}) for $g_{+}=6$, $g_{-}=2$,
    and $d_{-}=1.5$. The solid line corresponds to the half-period
    $T=0.5$, the short-dashed line (red online) to $T=1$, and the
    long-dashed line (blue online) to $T=1.5$. In these cases
    $f_{\mathrm{th}}=8,5,\,\textrm{and}\; 4$, respectively. The
    symbols (in color online) depict the numerical results that
    are obtained in the same way as in Fig.~6. }
\end{figure}


\begin{thebibliography}{99}

\bibitem{HM09}
P.~H\"{a}nggi and F.~Marchesoni, Rev. Mod. Phys. \textbf{81}, 387 (2009).

\bibitem{HMN}
P.~H\"{a}nggi, F.~Marchesoni, and F.~Nori, Ann.\ Phys.\ (Leipzig) \textbf{14},
51 (2005).
\bibitem{AH}
R.~D.~Astumian and P.~H\"{a}nggi, Phys.\ Today\ \textbf{55} (11), 33
(2002).
\bibitem{Rei}
P. Reimann and P. H\"anggi, Appl. Phys. A \textbf{75}, 169 (2002); P.~Reimann,
Phys.\ Rep.\ \textbf{361}, 57 (2002).

\bibitem{JAP}
F.~J\"{u}licher, A.~Ajdari, and J.~Prost, Rev.\ Mod.\ Phys.\ \textbf{69}, 1269
(1997).

\bibitem{KMWD}
K.W.~Kehr, K.~Mussawisade, T.~Wichmann, and W.~Dieterich, Phys.\ Rev.\ E\
\textbf{56}, R2351 (1997).
\bibitem{DLB}
I.~Der\'{e}nyi, C.~Lee, and A.-L.~Barab\'{a}si, Phys.\ Rev.\ Lett.\
\textbf{80}, 1473 (1998).
\bibitem{GM}
R.~Guantes and S.~Miret-Art\'{e}s, Phys.\ Rev.\ E\ \textbf{67}, 046212 (2003).
\bibitem{SGMH}
S.~Sengupta, R.~Guantes, S.~Miret-Art\'{e}s, and P.~H\"{a}nggi, Physica\ A\
\textbf{338}, 406 (2004).


\bibitem{BHK}
R. Bartussek, P. H\"anggi and J.G. Kissner, Europhys. Lett.  {\bf
28}, 459 (1994).

\bibitem{DHR}
C.~R.~Doering, W.~Horsthemke, and J.~Riordan, Phys.\ Rev.\ Lett.\
\textbf{72}, 2984 (1994).
\bibitem{JKH}
P.~Jung, J.~G.~Kissner, and P.~H\"{a}nggi, Phys.\ Rev.\ Lett.\ \textbf{76},
3436 (1996).
\bibitem{Mat}
J.~L.~Mateos, Phys.\ Rev.\ Lett.\ \textbf{84}, 258 (2000).
\bibitem{SAM}
R.~Salgado-Garc\'{i}a, M.~Aldana, and G.~Mart\'{i}nez-Mekler, Phys.\ Rev.\
Lett.\ \textbf{96}, 134101 (2006).
\bibitem{KSP}
A.~Kenfack, S.~M.~Sweetnam, and A.~K.~Pattanayak, Phys.\ Rev.\ E\ \textbf{75},
056215 (2007).

\bibitem{Linke} Special issue on \textit{Ratchets and Brownian
motors: Basics, experiments and applications}, edited by H.~Linke [Appl.\
Phys.\ A: Mater. Sci. Process. \textbf{75}, 167 (2002)].

\bibitem{HL}
T.~Harms and R.~Lipowsky, Phys.\ Rev.\ Lett.\ \textbf{79}, 2895 (1997).
\bibitem{March}
F.~Marchesoni, Phys.\ Rev.\ E\ \textbf{56}, 2492 (1997).

\bibitem{PASF}
M.~N.~Popescu, C.~M.~Arizmendi, A.~L.~Salas-Brito, and F.~Family, Phys.\ Rev.\
Lett.\ \textbf{85}, 3321 (2000).
\bibitem{AFS}
C.~M.~Arizmendi, F.~Family, and A.~L.~Salas-Brito, Phys.\ Rev.\ E\ \textbf{63},
061104 (2001).
\bibitem{GLZH}
L.~Gao, X.~Luo, S.~Zhu, and B.~Hu, Phys.\ Rev.\ E\ \textbf{67}, 062104 (2003).
\bibitem{ZLAF}
D.~G.~Zarlenga, H.~A.~Larrondo, C.~M.~Arizmendi, and F.~Family, Phys.\ Rev.\ E\
\textbf{75}, 051101 (2007).

\bibitem{DDH}
S.~I.~Denisov, E.~S.~Denisova, and P.~H\"{a}nggi, Phys.\ Rev.\ E\ \textbf{71},
016104 (2005).

\bibitem{CPAMH}
D. Cubero, J. Casado Pascual, A. Alvarez, M. Morillo, and P. H\"anggi, Acta
Phys. Pol. B \textbf {37}, 1467 (2006).

\bibitem{DKDH1}
S.~I.~Denisov, M.~Kostur, E.~S.~Denisova, and P.~H\"{a}nggi, Phys.\ Rev.\ E\
\textbf{75}, 061123 (2007).
\bibitem{DKDH2}
S.~I.~Denisov, M.~Kostur, E.~S.~Denisova, and P.~H\"{a}nggi, Phys.\ Rev.\ E\
\textbf{76}, 031101 (2007).

\end{thebibliography}
\end{document}